# Extending Network Intrusion Detection with Enhanced Particle Swarm Optimization Techniques

Surasit Songma[1], Watcharakorn Netharn[2*] and Siriluck Lorpunmanee[3]

[1]Department of Cyber Security, Faculty of Science and Technology, Suan Dusit University, Bangkok, Thailand
[2]Department of Information Technology, Faculty of Science and Technology, Suan Dusit University, Bangkok, Thailand
[3]Department of Computer Science, Faculty of Science and Technology, Suan Dusit University, Bangkok, Thailand

## ABSTRACT

*The present research investigates how to improve Network Intrusion Detection Systems (NIDS) by combining Machine Learning (ML) and Deep Learning (DL) techniques, addressing the growing challenge of cybersecurity threats. A thorough process for data preparation, comprising activities like cleaning, normalization, and segmentation into training and testing sets, lays the framework for model training and evaluation. The study uses the CSE-CIC-IDS 2018 and LITNET-2020 datasets to compare ML methods (Decision Trees, Random Forest, XGBoost) and DL models (CNNs, RNNs, DNNs, MLP) against key performance metrics (Accuracy, Precision, Recall, and F1-Score). The Decision Tree model performed better across all measures after being fine-tuned with Enhanced Particle Swarm Optimization (EPSO), demonstrating the model's ability to detect network breaches effectively. The findings highlight EPSO's importance in improving ML classifiers for cybersecurity, proposing a strong framework for NIDS with high precision and dependability. This extensive analysis not only contributes to the cybersecurity arena by providing a road to robust intrusion detection solutions, but it also proposes future approaches for improving ML models to combat the changing landscape of network threats.*

## 1. INTRODUCTION

Network security is one of the significant challenges that network administrators and owners face, particularly given the growing number and complexity of attacks. Due to the rapid increase in those issues, various protection measures and methods must be developed. Network Intrusion Detection Systems (NIDS) scan and analyze network traffic to identify assaults and alert network administrators[1][2][3].

In recent years, the growth of digital communication networks has resulted in an unprecedented volume of data, requiring improved machine learning (ML) algorithms for successful analysis and categorization [4]. The CSE-CIC-IDS 2018 and LITNET-2020 datasets, a representative example of such extensive datasets, present challenges in multi-class classification tasks[3][4][5][6][7]. The objective of this paper is to conduct a thorough investigation into the performance of several ML classifiers, such as Random Forest (RF), Decision Tree (DT), Extreme Gradient Boosting (XGBoost)[3][6][8],Convolutional Neural Networks (CNNs)[9], Recurrent Neural Networks (RNNs), Deep Neural Networks (DNNs)[11],and Multilayer Perceptron (MLP)[6][12][13]on the CSE-CIC-IDS 2018 and LITNET-2020 datasets. As a result, the best model is selected for fine-tuning with Enhanced Particle Swarm Optimization (EPSO)[14][15].

The selection of classifiers for this study reflects a diverse range of approaches, from traditional clustering methods to state-of-the-art ensemble models and neural network architectures. In contrast, RF, DT, and XGBoost, renowned for their robustness and accuracy, are investigated as representative ensemble methods. Additionally, the study explores the effectiveness of Deep learning (DL) models, including CNNs, RNNs, DNNs, and MLP, which are known for their capability to automatically extract complex features from high-dimensional data.

The motivation behind this research lies in the necessity to discern the strengths and limitations of diverse ML approaches when confronted with the challenges posed by the CSE-CIC-IDS 2018 and LITNET-2020 datasets. Understanding the relative performance of these classifiers is pivotal for guiding the selection of models in real-world applications, particularly in domains where multi-class classification plays a crucial role, such as network security and anomaly detection.Through a rigorous evaluation based on accuracy, precision, recall, and F1-Score metrics[6][11][16]., this research aims to provide insights into the suitability of each classifier for the CSE-CIC-IDS 2018 and LITNET-2020 datasets, offering valuable guidance for practitioners and researchers alike in choosing the most effective approach for similar complex classification tasks.

The goal of this study was to find the most efficient classifier using preprocessing methodologies, which convert into widely used ML and DL approaches for intrusion detection systems(IDS). We investigated popular and high-performing classification techniques such as XGBoost, DT, RF,CNNs, RNNs, DNNs, and MLP. The performance evaluation was multidimensional, focusing on four essential metrics: accuracy, precision, recall, and F1-Score. Accuracy is a critical metric of a model's effectiveness in classification tasks, expressing the proportion of correct predictions to total predictions made. A model with great accuracy can foresee outcomes that are consistent with real-world observations.The primary contributions of this work can be described in full detail as the following:

- Using the CSE-CIC-IDS-2018 and LITNET-2020 datasets, which are among the most recent and include a wide range of incursion types, establishing them as standards.
- On these datasets, researchers evaluated well-known ML classifiers with DL models, that is, a combination had never been substantially examined.
- Implementation of a multi-class classification approach to assess model efficiency thoroughly.
- Verification of model adequacy across both datasets to ensure optimal fit and applicability to a variety of data scenarios.
- Model performance is improved by using EPSO for more accuracy and finer parameter selection.
- Researchers evaluated the performance of ML classifiers that they had previously classified as effective[3].
- The analysis of full datasets took into account Big Data, setting this effort apart from others that frequently use a random subset of data for research.

The structure of the paper is as follows Section 2 reviews relevant previous studies, Section 3 describes the proposed research methodology, Section 4 discusses the datasets chosen for analysis, Section 5 details the experimental setup, Section 6 presents the results and discussions of the experiments, and Section 7 concludes the paper with a summary of the findings and recommendations for future work

## 2. RELATED WORKS

Research into multi-class classification using ML classifiers has yielded extensive discourse, delving into the comparative strengths and weaknesses of various algorithms applied to a range of datasets. Ensemble strategies, notably RF, DT, and XGBoost, have risen as formidable contenders in the classification arena, attributed to their proficiency in curbing overfitting and enhancing predictive precision[17]. Empirical studies have corroborated the superiority of these methods in diverse sectors, warranting their inclusion for an in-depth evaluation against the CSE-CIC-IDS 2018 and LITNET-2020 datasets within this study[18][19].

The proliferation of DL has amplified the deployment of neural network models, especially CNNs, RNNs, DNNs, and MLP in sophisticated classification scenarios[4]. CNNs are lauded for their inherent ability to autonomously distill hierarchical features from structured inputs, proving formidable in image and sequence processing tasks. MLP are celebrated for their aptitude in deciphering intricate data interrelations and securing roles across various disciplines. RNNs are distinguished by their specialized design to process sequential information, adeptly capturing temporal sequences. The established adaptability of these neural network models in past research underpins their selection for this study's analytical assessment using the CSE-CIC-IDS 2018 and LIT-NET-2020 datasets.

The literature presents extensive analyses of individual classifiers, ensemble methods, and neural networks on the intricate CSE-CIC-IDS 2018 and LITNET-2020 datasets. This study endeavors to fill this void by thoroughly comparing these classifiers, taking into account their unique attributes and appropriateness for complex multi-class classification tasks. Our goal is to improve the existing knowledge base on the ML model, particularly inside the complex frameworks of digital communication networks.

Simultaneously, there's a burgeoning interest in developing hybrid models that synergize the descriptive power of clustering techniques with the robust predictive abilities of advanced ensemble methods and neural networks. By exploring hybrid models on the CSE-CIC-IDS 2018 and LITNET-2020 datasets, this research may uncover novel insights into effective strategies for managing the sophisticated challenges of multi-class classification. This is particularly pertinent in the field of network security and anomaly detection, where precision in classification is paramount for identifying malicious threats. Against the backdrop of rapid advancements in ML, this investigation will also consider the latest optimization algorithms, regularization methods, and architectural innovations, ensuring a state-of-the-art approach in classifier evaluation.

A review of key documents and research publications revealed that various studies used ML techniques in conjunction with the CSE-CIC-IDS 2018 and LIT-NET-2020 datasets to detect intrusions. This can be illustrated as the following:

- W. Chimphlee and S. Chimphlee,2024 [8]demonstrated that the HO-XGB algorithm, with tuned hyperparameters such as learning_rate, subsample, and max_depth, outperforms other ML in identifying network intrusions on the CSE-CIC-IDS-2018 dataset.
- S. Songma, T. Sathuphan, and T. Pamutha, 2023 [3]improved intrusion detection on the CSE-CIC-IDS-2018 dataset by utilizing data preprocessing, PCA, and RF for feature reduction. They discovered that the XGBoost, DT, and RF models worked best, with PCA and RF improving both performance and efficiency.

- S. Chimphlee and W. Chimphlee, 2023 [6] determined that the MLP algorithm excelled in detecting network intrusions on the CSE-CIC-IDS-2018 dataset. After data preprocessing and feature selection with RF, MLP outperformed other algorithms, including LR, KNN, CART, Bayes, RF, and XGBoost.
- S. Nath, D. Pal, and S. Saha, 2023 [20] proposed a "honeyed framework" for detecting illegal actions in IoT environments, including DoS assaults. This system effectively identifies and mitigates vulnerabilities, as demonstrated by analysis of the IoT-23, NetML-2020, and LITNET-2020 datasets.
- A. A. Awad, A. F. Ali, and T. Gaber, 2023 [12] created an ILSTM model for intrusion detection that was optimized utilizing CBOA and PSO. On the LITNET-2020 and NSL-KDD datasets, this model outperformed standard LSTM and other deep learning models in terms of accuracy and precision.
- V. Hnamte and J. Hussain, 2023 [11] showed that the DCNN model, using deep CNNs with GPU acceleration, achieved 99.79% to 100% accuracy in threat detection. This model was tested on large datasets like ISCX-IDS 2012, DDoS, CICIDS2017, and CSE-CIC-IDS-2018, demonstrating its high efficacy in IDS performance.
- V. Bulavas, V. Marcinkevičius, and J. Rumiński, 2021 [21] utilized SMOTE to address class imbalance in network intrusion detection by upsampling rare classes. They discovered that DT ensembles (Adaboost, RF, Gradient Boosting Classifier) outperformed CIC-IDS2017, CSE-CIC-IDS2018, and LITNET-2020 datasets.
- H. Al-Zoubi and S. Altaamneh, 2022 [22] developed a CCSA-based feature selection technique for NIDS that increased accuracy, detection rate, and precision, and reduced false alarms. On the LITNET-2020 dataset, CCSA outperformed standard classifiers such as KNN, DT, RF, SVM, MLP, and LSTM.
- L. Yang, J. Li, L. Yin, Z. Sun, Y. Zhao, and Z. Li, 2020 [16] developed a CDBN-based technique with "SamSelect" and SCAE to enhance WLAN security. This method improves real-time attack detection through intrusion detection, data balance, and dimensionality reduction, achieving high speed and accuracy on AWID and LITNET-2020 datasets.
- V. Dutta, M. Choraś, M. Pawlicki, and R. Kozik, 2020 [4] For the detection of network anomalies in IoT contexts, an ensemble method was created employing DNNs, LSTM, logistic regression, and DSAE. This strategy, which uses stacking ensemble learning, outperformed traditional methods on the IoT-23, LITNET-2020, and NetML-2020 datasets.

As a result, previous researchers have investigated different common MLalgorithms for intrusion detection

## 3. PROPOSED METHODOLOGY

We gained an understanding of the present difficulties and proposed solutions by examining relevant literature and scholarly papers. This informed understanding has been simplified into a complete overview of ML and DL methods used in NIDS, as shown in Figure 1

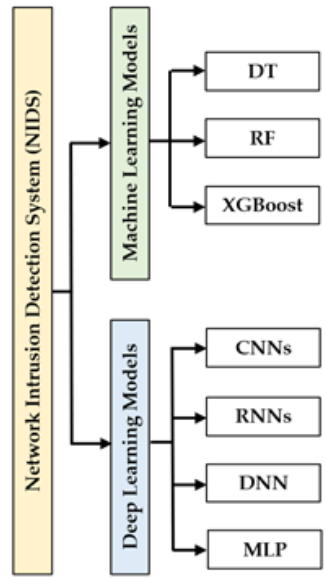

Figure 1. The methodology employs a taxonomy of machine and DL models in NIDS.

Figure 1. depicts a classification of ML and DL models that are often used in NIDS. It divides models into two major branches.

### 3.1. Machine Learning Models

This branch includes models that use ML approaches. This category includes the following[3].

- DT is a graph-like structure in which each internal node represents a "test" on an attribute, each branch reflects the test's conclusion, and each leaf node represents a class label[13][23][24].
- RF isan ensemble learning method that creates a huge number of DT during training and outputs the class that is the average of the classes in the individual tree[6][25][26].
- XGBoost Stands for eXtreme Gradient Boosting, which is an efficient and scalable version of the gradient boosting framework that improves its speed and performance[6][8][27].

### 3.2. Deep Learning Models

This section contains models built on DL architectures that are intended to build hierarchical representations of data[6][22][28][29][30].

- CNNs are a type of DNNs that is most typically used to analyze visual vision. They have layers that may extract features hierarchically using convolutional filters[11].
- RNNs are neural networks in which node connections create a directed graph that follows a temporal sequence. This structure allows the network to behave dynamically over time[31].
- DNNs is a neural network with a specific amount of complexity. It has numerous layers of nodes between the input and output layers, allowing it to represent complex nonlinear interactions[29].
- MLP A feedforward artificial neural network known as MLP has at least three layers of nodes that can detect non-linear correlations in data[13].

### 3.3. The Proposed Framework

Figure 1 could be used in an academic article or technical presentation to describe the implementation of NIDS using a variety of ML and DL techniques. Figure 2 depicts the conceptual model of the operational framework produced by the researcher. This framework builds on the researcher's earlier work, which focused on optimizing IDS utilizing three phase strategy [3], and the CSE-CIC-IDS-2018 and LITNET-2020 Datasetsand integrates aspects from that work.This flowchart from Figure 2 describes the typical process of building and evaluating ML and DL models. It outlines the following key phases.

- The method begins with the original datasets, which in this case are the CSE-CIC-IDS 2018 and LITNET-2020. These datasets provide raw data for training and testing the models.
- Data Preprocessing This is the first phase, in which raw data is prepared for modeling. This includes:
- Data Cleaning Delete or correct data that is erroneous, incomplete, irrelevant, duplicated, or incorrectly formatted.
- Exploratory Data Analysis Analyzing data to find patterns, relationships, or anomalies to guide further analysis.
- Encoding is the process of transforming categorical data into a numerical format.
- Normalization involves rescaling input data to a uniform, standard range.
- Data Splitting Divide the dataset into training and testing sets.
- Training Phase The preprocessed training dataset is supplied into a ML or DL algorithm. This method builds a model by learning from the data.
- Testing Phase The trained model is tested on a separate testing dataset. This stage is critical for evaluating the model's performance on new, previously unseen data.
- Evaluation Model This step entails evaluating the performance of ML/DL models using metrics like accuracy, precision, recall, and F1-Score. Based on these criteria, the best-performing model is chosen. The model's performance is assessed using the following metrics.
- Accuracy refers to the percentage of correct predictions made by the model.
- Precision is calculated as the number of real positive results divided by the total number of positive outcomes, including those that were incorrectly detected.
- Recall is the number of true positive results divided by the total number of samples that should have been positive.
- F1-Score is the harmonic mean of precision and recall, achieving a balance between the two criteria.
- Optimization Finally, the selected model undergoes optimization using Extended Particle Swarm Optimization (EPSO) to further refine the model's parameters and improve its performance.

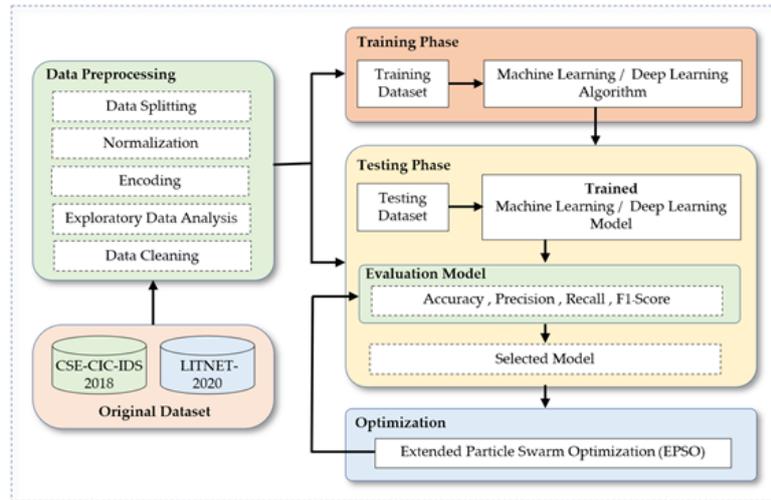

Figure 2. The proposed framework.

The flowchart presents a standard approach to ML model development, showcasing the importance of data preparation, model training, and model evaluation to ensure that the developed model performs accurately and effectively on the task it's designed for.

## 4. SELECTION OF DATASET

Recent advances in cybersecurity research have made new flow-based benchmark datasets available to the public, including CSE-CIC-IDS-2018 and LIT-NET-2020 [4][21].These datasets are critical to enhancing the efficacy of supervised learning systems in network intrusion detection. They provide the vital data necessary to properly train IDS, allowing these systems to identify a wide range of network threats with high accuracy and dependability. Using such datasets, IDS may be fine-tuned to detect various sorts of cyber threats, resulting in a strong defense against network intrusions.

### 4.1. CSE-CIC-IDS-2018 Datasets

CSE-CIC-IDS-2018 dataset is a collaboration between the Communications Security Establishment (CSE) and the Canadian Institute for Cybersecurity (CIC), and it is meant to boost intrusion detection research[3]. It has become the standard for evaluating IDSs [6][32]. This dataset was created to realistically depict the variety and complexity of cyber threats and attacks, and it includes a wide range of situations for analysis. Its usefulness stems from its capacity to replicate complicated network environments, allowing academics and practitioners to successfully test and enhance IDS technology. The information, which was collected over ten days and has eighty columns, comprises fifteen different types of assaults, such as FTP and SSH brute force, numerous DoS and DDoS attacks, web attacks, SQL injection, infiltration, and botnet activities, making it a comprehensive tool for cybersecurity research.

### 4.2. LITNET-2020 Datasets

The LITNET-2020 dataset represents a significant advancement in cybersecurity research, dramatically enhancing network intrusion detection across traditional computer networks, Wireless Sensor Networks (WSN), and the Internet of Things (IoT). Moving beyond the limitations of traditional network-intrusion benchmark datasets, which struggle to accurately reflect the complexity of modern network traffic and cyberattack patterns, LITNET-2020 introduces a variety of genuine network traffic scenarios as well as meticulously annotated examples of real attacks, avoiding the reliance on synthetic attacks commonly seen in sandboxed environments. The information systematically categorizes eighty-five particular network flow variables and defines twelve distinct types of attacks, covering a wide range of network behaviors from regular operations to cyber threats. Built on a solid infrastructure that uses Cisco routers and FortiGate (FG-1500D) firewalls to generate NetFlow data, as well as a sophisticated collector system for data management, storage, and analysis, LITNET-2020 provides a detailed and authentic resource for crafting and improving security measures to combat the dynamic nature of network vulnerabilities and threats[5][20].

## 5. EXPERIMENTAL SETUP

This study was carried out on a 64-bit Windows 10 system equipped with an Intel® Xeon® Silver 4314 processor and 62 GB of DDR4 RAM. Due to the large volume of data, the Python 3.11 environment was utilized, which benefits from regular updates that improve language features, performance, and libraries. Numpy and Pandas were utilized for data handling and preprocessing, and Scikit Learn was used for model training, evaluation, and metric assessment. Data visualization was done with Seaborn and Matplotlib. Further information can be found in the subsections below.

### 5.1. Import Full Dataset

First, ensure that the CSE-CIC-IDS-2018 and LITNET-2020 dataset is produced in a manner that allows for easy importing into your data analysis tool, such as CSV, Excel, JSON, or via a database linkage. Then, load the dataset using relevant libraries or tools from your chosen computer language, such as Python with pandas, R, or SQL, which may include file reading, database connection, or API interaction. After successfully importing the dataset, make sure it is kept in a variable or structured data format for future analysis. Tables 1-2 provide more detailed information.

Table 1. CSE-CIC-IDS-2018 Network Traffic Distribution by Labelled Attacks Type and Ratio.

| Attacks Type | Amount of Data | Ratio (%) |
|---|---|---|
| Benign | 13,484,708 | 83.0697 |
| DDoS | 1,263,933 | 7.7862 |
| DoS | 654,300 | 4.0307 |
| Brute Force | 380,949 | 2.3468 |
| Botnet | 286,191 | 1.7630 |
| Infiltration | 161,934 | 0.9976 |
| Web attacks | 987 | 0.0061 |
| **Total** | **16,233,002** | **100.0000** |

Table 1 depicts the distribution of data types in a dataset used for NIDS with 16,233,002 instances. It divides the data into benign and attack categories, with benign traffic accounting for 83.07% (13,484,708 instances)[21]. DDoS attacks are the most common of the near-perfect s, accounting for 7.79% of the dataset, followed by DoS, brute force, botnet, infiltration, and web attacks, which account for the remainder. The data show that normal network traffic is much more common than harmful attempts, offering a thorough perspective for analyzing and upgrading network security measures.

Table 2. LITNET-2020 Network Traffic Distribution by Labelled Attacks Type and Ratio.

| Attacks Type | Amount of Data | Ratio (%) |
|---|---|---|
| none | 36,423,860 | 91.9709 |
| Smurf | 118,958 | 0.3004 |
| ICMP-flood | 23,256 | 0.0587 |
| UDP-flood | 93,583 | 0.2363 |
| TCP SYN-flood | 1,580,016 | 3.9896 |
| HTTP-flood | 22,959 | 0.0580 |
| LAND attack | 52,417 | 0.1324 |
| Blaster Worm | 24,291 | 0.0613 |
| Code Red Worm | 1,255,702 | 3.1707 |
| Spam bot's detection | 747 | 0.0019 |
| Reaper Worm | 1,176 | 0.0030 |
| Scanning/Spread | 6,232 | 0.0157 |
| Packet fragmentation attack | 477 | 0.0012 |
| **Total** | **39,603,674** | **100.0000** |

Table 2 categorizes numerous types of network attacks in a dataset of 39,603,674 entries, indicating their amounts and percentages of the total. The vast majority of the data (91.9709%) is classed as 'none', suggesting typical, non-attack traffic[21]. Among the attack kinds, the most common is 'TCP SYN-flood', which accounts for 3.9896% of the data, followed by 'Code Red Worm' attacks at 3.1707%. Other prominent assault types include 'Smurf', 'UDP-flood', and 'ICMP-flood', among others, each accounting for a minor fraction of the dataset. The variety and proportions of these attacks illustrate the dataset's composition, demonstrating both benign and malicious traffic, as well as extensive enumeration for analytical and modeling purposes.

### 5.2. Understanding the Features and Describing them

Begin by evaluating the dataset's structure using commands such as df.head() in Python with pandas or SELECT TOP 5 * in SQL to study the first data rows. It is critical to comprehend each column's relevance and purpose, which may require studying the dataset's data dictionary or associated documentation. Examine the data types used throughout characteristics, such as numerical, categorical, and date-time. To understand crucial statistical factors, construct summary statistics for numerical features such as mean, median, and standard deviation. Use visualization tools likehistograms, scatter plots, and box plots to investigate numerical feature distributions. Perform frequency counts on categorical features and show the distributions with bar plots or pie charts. Finally, evaluate and address missing values in each feature using imputation or removal to prepare the data for future analysis.

### 5.3. Count Unique Records

After familiarizing yourself with the dataset's features, count the unique records to see if there are any duplicates, which are indicated by rows that have identical values across all columns. Use a command or function to count unique entries, using either a primary key or a precise combination of columns that uniquely identify each item. This technique will disclose the dataset's total number of unique records, which will aid in ensuring data integrity and accuracy for future analysis.

### 5.4. Clean NaN, Missing, and Infinite values from the Dataset

Identify columns in your dataset with NaN (null) values, missing data, or infinite values (positive or negative infinity). Choose how to handle missing or infinite data, such as imputation (filling gaps with mean, median, or mode), eliminating rows with missing values, or using more advanced imputation methods. For dealing with infinite values, consider replacing them with appropriate numbers or eliminating the rows containing these values, especially if they are deemed outliers. This method ensures that the dataset is appropriately prepared, with consistent and useful data for subsequent analysis. After cleansing the dataset of NaN (Null), missing, and infinite values, it was revealed that there were no such values, suggesting that the dataset was complete and consistent, with no missing or undefined entries.

## 5.5. Clean Data by Removing Duplicate Rows

Identify and discover duplicate rows in the dataset using a duplicate-detecting command or function, concentrating on specific columns as appropriate. Depending on the objectives of your research, decide whether to keep the first occurrence of duplication while removing the others, or vice versa. To eliminate duplicate entries, use the appropriate approach; for example, in Python with pandas, use the df.drop_duplicates() function, but in SQL, use the DELETE statement with a WHERE clause. This method guarantees that the dataset is devoid of duplicate entries, leaving it clean and ready for future analysis. After successfully cleaning the data by deleting duplicate rows, Tables 3 and 4 provide extensive information.

Table 3. Comparative Analysis of Data Before and After Duplicate Removal in the CSE-CIC-IDS-2018 Dataset by Attack Type.

| Attacks Type | Original | | Remove Duplicate | |
|---|---|---|---|---|
| | Amount of Data | Ratio(%) | Amount of Data | Ratio(%) |
| Benign | 13,484,708 | 83.0697 | 10,666,030 | 88.7517 |
| DDoS | 1,263,933 | 7.7862 | 775,955 | 6.4567 |
| DoS | 654,300 | 4.0307 | 196,568 | 1.6356 |
| Brute Force | 380,949 | 2.3468 | 94,101 | 0.7830 |
| Botnet | 286,191 | 1.7630 | 144,535 | 1.2027 |
| Infiltration | 161,934 | 0.9976 | 139,775 | 1.1631 |
| Web attacks | 987 | 0.0061 | 867 | 0.0072 |
| **Total** | **16,233,002** | **100.0000** | **12,017,831** | **100.0000** |

Table 3 compares the original and deduplicated network attack data from the CSE-CIC-IDS-2018 dataset. It demonstrates the effectiveness of deleting duplicate entries across a wide range of threats, including benign traffic and malicious activity such as DDoS, DoS, Brute Force, Botnet, Infiltration, and Web attacks. Initially, the collection had 16,233,002 records, with benign traffic accounting for 83.07% of the data. After deduplication, the total data was decreased to 12,017,831 items, with benign traffic increasing to 88.75%. Notably, all types of assaults witnessed a decrease in data volume, with DDoS, DoS, and Brute Force attacks experiencing the greatest decreases, indicating that these categories contained a considerable quantity of duplicate data. The approach successfully reduced dataset size while changing the ratio percentages of each attack type, emphasizing the necessity of data cleaning in preparing accurate and efficient datasets for analysis.

Table 4. Comparative Analysis of Data Before and After Duplicate Removal in the LITNET-2020 Dataset.

| Attacks Type | Original | | Remove Duplicate | |
| --- | --- | --- | --- | --- |
| | Amount of Data | Ratio(%) | Amount of Data | Ratio(%) |
| none | 36,423,860 | 91.9709 | 32,087,765 | 91.1676 |
| Smurf | 118,958 | 0.3004 | 59,479 | 0.1690 |
| ICMP-flood | 23,256 | 0.0587 | 11,628 | 0.0330 |
| UDP-flood | 93,583 | 0.2363 | 93,583 | 0.2659 |
| TCP SYN-flood | 1,580,016 | 3.9896 | 1,580,016 | 4.4891 |
| HTTP-flood | 22,959 | 0.0580 | 22,959 | 0.0652 |
| LAND attack | 52,417 | 0.1324 | 52,417 | 0.1489 |
| Blaster Worm | 24,291 | 0.0613 | 24,291 | 0.0690 |
| Code Red Worm | 1,255,702 | 3.1707 | 1,255,702 | 3.5677 |
| Spam bot's detection | 747 | 0.0019 | 747 | 0.0021 |
| Reaper Worm | 1,176 | 0.0030 | 1,176 | 0.0033 |
| Scanning/Spread | 6,232 | 0.0157 | 6,232 | 0.0177 |
| Packet fragmentation attack | 477 | 0.0012 | 477 | 0.0014 |
| **Total** | **39,603,674** | **100.0000** | **35,196,472** | **100.0000** |

Table 4. depicts the impact of deleting duplicate records from the LITNET-2020 dataset, which originally had 39,603,674 data points. The 'none' category, which indicates no attack, accounted for around 91.97% of the original dataset and reduced slightly to 91.17% after duplicates were deleted. Attack categories such as 'Smurf' and 'ICMP-flood' have much lower numbers after duplication removal, affecting their proportions within the dataset. Other categories, such as 'UDP-flood' and 'TCP SYN-flood', preserved their previous numbers, indicating no duplication. The elimination process changed the ratios for each assault type, resulting in a more realistic depiction of the dataset's composition. Following cleanup, the dataset now contains 35,196,472 records, ensuring that subsequent studies are based on unique and non-repetitive data, which is critical for reliable cybersecurity research and modeling.

### 5.6. Consolidating and Removing Unnecessary Features

Consolidating and removing unnecessary features includes simplifying a dataset by combining similar qualities and discarding those that are irrelevant to the model's performance. This procedure improves data quality, reduces model complexity, and increases computing performance, making the dataset more suitable for predictive analytics and ML applications

### 5.6.1. Streamlining Features of the CSE-CIC-IDS-2018 Dataset

Following an initial review of the CSE-CIC-IDS-2018 dataset, which originally contained 16,233,002 entries across 80 attributes, duplicate record elimination reduced the dataset to 12,017,831 items. Furthermore, eight attributes with constant zero values for each record were removed, including "Bwd PSH Flags," "Bwd URG Flags," "FwdByts/b Avg," "FwdPkts/b Avg," "FwdBlk Rate Avg," "BwdByts/b Avg," "BwdPkts/b Avg," and "BwdBlk Rate Avg". The "Timestamp" feature was also eliminated to avoid biasing the model toward attack prediction or detection. Further data cleaning included the removal of entries with "NaN" values, as well as two attributes with constant infinite values, "Flow Byts/s" and "Flow Pkts/s". With these changes, the improved dataset now contains 12,017,831 records and 69 features, making it better suited to classification tasks and additional analytical research.

### 5.6.2. Streamlining Features of the LITNET-2020 Dataset

#### 5.6.2.1. Convert Data with Object Type to Numeric for Algorithm Preparation

To efficiently apply ML algorithms to your dataset, identify columns that contain object or non-numeric data types, which are frequently filled with text descriptions of various network properties. These contain categorical variables like'sa', 'da', 'pr', '_flag1', through '_flag6', 'nh', 'nhb', 'ismc', 'odmc', 'idmc', 'osmc','mpls1' to'mpls10', 'ra', 'eng', 'tr', 'icmp_dst_ip_b', 'icmp_src_ip', 'udp_'  'tcp_f_s', 'tcp_f_n_a' to 'tcp_f_n_u', 'tcp_dst_p', 'tcp_src_dst_f_s', 'tcp_src_tftp', 'tcp_src_kerb', 'tcp_src_rpc', 'tcp_dst_p_src', 'smtp_dst', 'udp_p_r_range', 'p_range_dst', 'udp_src_p_0', among others. It is critical to convert these variables to a numeric format using methods such as label encoding or custom mapping. This transformation is necessary to ensure that the dataset fits the numerical criteria of ML and statistical models, enabling for the extraction of relevant, data-driven insights and analysis.

#### 5.6.2.2. Merge Features 1-6 to Start Timestamp and Features 7-12 to End Timestamp

Improve your dataset's analytical capabilities by identifying and combining attributes that capture an event's start and end times. Start times range from 'ts_year' to 'ts_second', whereas end times range from 'te_year' to 'te_second'. Merge these data points into two new columns,'stimestamp' for the event's start and 'etimestamp' for its end, by aggregating the date and time values from each connected feature set. Ensure that these additional columns are in datetime format to allow for time-based analysis, which will improve the practicality and precision of any temporal research while also streamlining the overall dataset structure.

#### 5.6.2.3. Remove Feature Names with Values of Zero

Following an initial review of the LITNET-2020 dataset, 39,603,674 records spanning 85 different features were identified. The removal of duplicate entries lowered the total number of records to 35,196,472. The dataset was further streamlined by combining features 1-6 into a single 'stimestamp' feature and features 7-12 into 'etimestamp', decreasing the feature count to 75. Subsequent statistical analysis revealed that fea-tures such as 'fwd', 'opkt', 'obyt', 'smk', 'dmk', 'dtos', '_dir', 'nh', 'nhb','svln', 'dvln', ' ismc', 'odmc', 'idmc', 'osmc', 'mpls1', 'mpls2','mpls3', 'mpls4', 'mpls5', 'mpls6', 'mpls7', 'mpls8', 'mpls9' , 'mpls10','cl', 'sl', 'al', 'ra', 'eng', 'tr' had zero variance and were there-fore removed, leaving 44 features in total. Furthermore, the 'ID' and 'attack_a' attrib-utes were removed because they were not relevant to further investigation. This updated dataset now has 35,196,472 records and 42 features, making it suitable for more advanced processing.

### 5.7. Convert Attack Type with Multiple Class

To prepare your dataset for ML, use one-hot encoding to convert each category within a categorical variable into several class columns, then label encoding to assign a unique integer identification to each. Identify the column that contains information about attack kinds. Multiple categorization requires that the attacktype categories remain intact, protecting the integrity of all unique classifications. This approach ensures that the data is represented comprehensively, allowing for more detailed analysis and modeling.

### 5.8. Perform Statistical Analysis for Min, Max, STD, and Mean for Each Feature

To gain insights into the numerical features of your dataset, calculate descriptive statistics for your dataset's numerical properties, such as the minimum, maximum, standard deviation, and mean. Using built-in tools, like df.describe() in pandas, provides a quick overview of these fundamental statistical indicators. This stage is crucial for understanding your data's distribution, variability, and core patterns since it lays the groundwork for future data exploration and modeling activities.

## 5.9. Normalize Min-Max features

Min-max scaling is an effective method for standardizing numerical features in ML algorithms that need normalized input, such as neural networks and support vector machines. This procedure adjusts the features to a certain range, usually between 0 and 1, ensuring that their magnitudes are equal and preventing any single feature from excessively influencing the learning process. The min-max scaling is accomplished by using the mathematical equation shown as (1), where X is an original value and X′ is the normalized value[3][6][33][34].

$$X' = \frac{(X - X_{min})}{(X_{max} - X_{min})} \quad (1)$$

## 5.10. Split a Dataset into Training and Testing Sets

Once the data has been prepared, it is ready for model testing, which involves splitting it into experimental and test datasets in an 80:20 ratio. This separation ensures that a significant percentage of the data is used for training and fine-tuning the model, while another, smaller sample is saved for testing its performance under situations similar to real-world applications[35]. Detailed information about this distribution is properly organized in Table 5-6.

Table 5. Training and Testing Dataset Distribution by Attack Type for the CSE-CIC-IDS-2018 Dataset.

| Attacks Type | Training | | Testing | |
| --- | --- | --- | --- | --- |
| | Amount of Data | Ratio(%) | Amount of Data | Ratio(%) |
| Benign | 8,532,824 | 88.7517 | 2,133,206 | 88.7517 |
| DDoS | 620,764 | 6.4567 | 155,191 | 6.4567 |
| DoS | 157,254 | 1.6356 | 39,314 | 1.6356 |
| Brute Force | 75,281 | 0.7830 | 18,820 | 0.7830 |
| Botnet | 115,628 | 1.2027 | 28,907 | 1.2027 |
| Infiltration | 111,820 | 1.1631 | 27,955 | 1.1631 |
| Web attacks | 694 | 0.0072 | 173 | 0.0072 |
| **Total** | **9,614,265** | **100.0000** | **2,403,566** | **100.0000** |

Table 6. Training and Testing Dataset Distribution by Attack Type for the LITNET-2020 Dataset.

| Attacks Type | Training | | Testing | |
| --- | --- | --- | --- | --- |
| | Amount of Data | Ratio(%) | Amount of Data | Ratio(%) |
| none | 25,670,212 | 91.1676 | 6,417,553 | 91.1676 |
| Smurf | 47,583 | 0.1690 | 11,896 | 0.1690 |
| ICMP-flood | 9,302 | 0.0330 | 2,326 | 0.0330 |
| UDP-flood | 74,866 | 0.2659 | 18,717 | 0.2659 |
| TCP SYN-flood | 1,264,013 | 4.4891 | 316,003 | 4.4891 |
| HTTP-flood | 18,367 | 0.0652 | 4,592 | 0.0652 |
| LAND attack | 41,934 | 0.1489 | 10,483 | 0.1489 |
| Blaster Worm | 19,433 | 0.0690 | 4,858 | 0.0690 |
| Code Red Worm | 1,004,562 | 3.5677 | 251,140 | 3.5677 |
| Spam bot's detection | 598 | 0.0021 | 149 | 0.0021 |

| Reaper Worm | 941 | 0.0033 | 235 | 0.0033 |
| Scanning/Spread | 4,986 | 0.0177 | 1,246 | 0.0177 |
| Packet fragmentation attack | 382 | 0.0014 | 95 | 0.0014 |
| **Total** | **28,157,178** | **100.0000** | **7,039,294** | **100.0000** |

### 5.11. Classification Model

Classification divides data into several groups, which is especially useful in IDS for distinguishing between benign and malicious network activity. It can be classified into two types binary classification, which deals with two distinct classes, and multi-class classification, which incorporates multiple classes. The addition of classes increases computational demand and time, which may reduce algorithm efficiency. During classification, data is examined to determine whether it is normal or abnormal. This method preserves existing data structures while allowing for the inclusion of new data points. Classification aids in the detection of unexpected patterns and irregularities, while it is most typically used to identify situations of misuse. Three ML and four DL models were utilized in this investigation.

ML models are divided into two categories standard ML approaches and DL models. In the traditional category, it mentions DT, which uses a tree-like model for decision-making, RF, an ensemble method that combines multiple DT for improved accuracy, and XGBoost, a highly efficient gradient boosting implementation that aims to optimize speed and performance. In the realm of DL, it highlights CNNs, which are especially effective for visual data analysis through hierarchical feature extraction, RNNs, designed to handle sequential data with their ability to model time-dependent information, DNNs, which are distinguished by their deep structure capable of modeling complex patterns, and MLP, a basic form.

### 5.12. Evaluation Model

The study assesses an intrusion detection system using measures such as accuracy, precision, recall, and F1-Score[6][11][16]. Accuracy is an important indication of anML model's performance, particularly for classification tasks, because it calculates the proportion of right predictions out of the complete dataset. A model with high accuracy may dependably predict outcomes that match the actual observed events.A confusion matrix is a specific table structure that allows for the presentation of an algorithm's performance, which is frequently supervised learning [33]. Each row of the matrix represents examples in an actual class, but each column represents instances in a predicted class, or vice versa. Here's a simple definition of the terms.

- True Positives (TP) These are instances where the model properly anticipated a positive outcome. It signifies that the actual class was positive, and the model expected a positive outcome.
- True Negatives (TN) These are the instances correctly predicted as negative by the model. It means that the actual class was negative, and the model also predicted it as negative.
- False Positives (FP) These are situations where the model mistakenly projected a positive outcome, also known as Type I errors. It signifies that the actual class was negative, while the model expected a positive outcome.
- False Negatives (FN) These are situations where the model mistakenly anticipated a negative value, also known as Type II error. It means that while the actual class was positive, the model projected a negative outcome.

#### 5.12.1. Accuracy

This metric assesses the model's overall correctness and is calculated as the ratio of correct predictions to total predictions made. It is acceptable when the class distribution is similar. The formula is[36].

$$Accuracy = \frac{(TP + TN)}{(TP + TN + FP + FN)} \qquad (2)$$

#### 5.12.2. Precision

Precision measures a model's ability to correctly identify only relevant instances. In other words, it represents the ratio of true positive forecasts to overall positive predictions. Here's the formula[36].

$$Precision = \frac{(TP)}{(TP + FP)} \qquad (3)$$

#### 5.12.3. Recall

Recall calculates the fraction of true positives properly detected by the model. Here's the formula[36].

$$Recall = \frac{(TP)}{(TP + FN)} \qquad (4)$$

#### 5.12.4. F1-Score

F1-Score is the harmonic mean of precision and recall, resulting in a balance of the two. It is especially effective when precision and recall need to be balanced, as well as when the class distribution is unequal. Here's the formula[36].

$$F1 - Score = \frac{2 \times (Precision \times Recall)}{(Precision + Recall)} \qquad (5)$$

### 5.13. Particle Swarm Optimization

Kennedy and Eberhart developed Particle Swarm Optimization (PSO) in 1995, borrowing inspiration from natural processes like bird flocking and fish schooling[12]. This strategy is used to solve optimization problems ranging from simple single-objective concerns to sophisticated multi-objective scenarios. PSO functions without requiring the problem's gradient, making it especially useful for nonlinear tasks where gradients are not available. It is frequently used in domains such as engineering, economics, and computer science for a number of applications, including neural network training, function minimization, and product optimization[14][15][37].
The core mechanism of PSO is a set of potential solutions, known as particles, that navigate across a solution space. These particles move according to mathematical rules that govern their location and velocity, taking into account both their personal best positions and the best positions discovered by the swarm. This dual impact directs the entire swarm toward optimal solutions via iterative changes. PSO's capacity to iteratively search for superior solutions while exploiting the swarm's collective learning makes it an effective tool for optimization problems.

## 6. EXPERIMENTAL RESULT AND DISCUSSIONS

This section will go over and discuss the experimental results. Previously, we used two cybersecurity datasets the CSE-CIC-IDS-2018 and LITNET-2020 datasets, and prepared the data through a series of sub-steps such as Data Cleaning, Exploratory Data Analysis, Encoding, Normalization, and Data Splitting to guarantee it was ready for classification. We separated the data into two groups training and testing. During the Training Phase, we used three ML methods for data classification DT, RF, and XGBoost, which were motivated by previous research titled "Optimizing Intrusion Detection Systems Using a Three-Phase Strategy and the CSE-CIC-IDS-2018 Dataset." We also employed four prominent DL models CNNs, RNNs, DNNs, and MLP. Following training, we evaluated the models' performance in the Testing Phase using conventional measures such as accuracy, precision, recall, F1-Score, and aggregated average score. These metrics are used to assess the prediction effectiveness and dependability of each classifier, with higher scores indicating better performance. The goal is to find the most efficient model for further optimization using EPSO by fine-tuning the parameters to improve results.The classifiers' performance will be presented in detail through tables and figures, providing both visual and numerical insights into their effectiveness. We want to get significant insights into future advances in IDS through extensive research and discussion. The following table and picture summarize the outcomes of our intensive experimentation.

Table 7. Performance Metrics for ML Classifiers Using CSE-CIC-IDS-2018 Dataset.

| Classifiers | Accuracy | Precision | Recall | F1-Score |
|---|---|---|---|---|
| RF | 0.990513682 | 0.96084231 | 0.859125805 | 0.893265813 |
| DT | 0.996709474 | 0.974639471 | 0.975416011 | 0.975026588 |
| XGboost | 0.992428337 | 0.958616558 | 0.904445321 | 0.927667115 |
| CNNs | 0.887517154 | 0.787686741 | 0.887517178 | 0.834627361 |
| DNNs | 0.887517154 | 0.787686741 | 0.887517178 | 0.834627361 |
| MLP | 0.887517178 | 0.787686741 | 0.887517178 | 0.834627361 |
| RNNs | 0.887746000 | 0.861307000 | 0.887746000 | 0.835293000 |

Table 7 compares the performance of different ML classifiers on the CSE-CIC-IDS-2018 dataset using metrics including Accuracy, Precision, Recall, and F1-Score. The RF classifier performs well, with high accuracy (0.990513682), and precision (0.96084231), but somewhat lower recall (0.859125805) and F1-Score (0.893265813), showing a reasonable balance between identifying true positives and minimizing false positives. The DT model beats others in terms of consistency across accuracy (0.974639471) and recall (0.975416011), with a near-perfect F1-Score (0.975026588), indicating that it captures the majority of positive cases without severe overprediction. XGboost also performs well in terms of accuracy (0.992428337) and precision-recall ratio, resulting in a high F1-Score (0.927667115). CNNs, DNNs, and MLP classifiers, on the other hand, outperform ensemble approaches in terms of accuracy (0.887517154), precision (0.787686741), recall (0.887517178), and F1-Score (0.834627361), with MLP showing slightly adjusted accuracy (0.887517178). This consistency shows that these networks may encounter difficulties because of the dataset's complexity or class imbalance. The RNNs model improves precision (0.861307000) while maintaining similar accuracy and F1-Score to the CNNs, DNNs, and MLP models, demonstrating a somewhat better but still limited competence in processing sequential data within this unique dataset. Overall, while ensemble models such as RF, DT, and XGboost perform better than neural net-work-based models (CNNs, DNNs, MLP, and RNNs) on the CSE-CIC-IDS-2018 dataset, highlighting the importance of model selection based on dataset characteristics and intended use case for cybersecurity threat detection.

Table 8. Performance Metrics for ML Classifiers Using LITNET-2020 Dataset.

| Classifiers | Accuracy | Precision | Recall | F1-Score |
|---|---|---|---|---|
| RF | 1.000000000 | 1.000000000 | 1.000000000 | 1.000000000 |
| DT | 1.000000000 | 1.000000000 | 1.000000000 | 1.000000000 |
| XGboost | 1.000000000 | 1.000000000 | 1.000000000 | 1.000000000 |
| CNNs | 0.989821970 | 0.983374832 | 0.989821992 | 0.986295708 |
| DNNs | 1.000000000 | 1.000000000 | 1.000000000 | 1.000000000 |
| MLP | 1.000000000 | 1.000000000 | 1.000000000 | 1.000000000 |
| RNNs | 0.997327268 | 0.995994388 | 0.997327289 | 0.996437922 |

Table 8 compares ML classifiers' performance on the LITNET-2020 dataset using four essential metrics accuracy, precision, recall, and F1-Score. Most classifiers, including RF, DT, XGBoost, DNNs, and MLP, had perfect scores of 1.000000000 across all criteria, suggesting that they accurately predicted every case. The CNNs and RNNs classifiers performed well, but could not achieve total perfection. The CNNs had an accuracy of 0.989821970, precision of 0.983374832, recall of 0.989821992, and F1-Score of 0.986295708. The RNNs had slightly greater accuracy and recall at 0.997327268 and 0.997327289, respectively, with a precision of 0.995994388 and an F1-Score of 0.996437922. This table shows how well these classifiers can predict and categorize network traffic, with CNNs and RNNs performing somewhat lower but still outstandingly compared to the other models.

Table 9. Comparative Performance of ML Classifiers on LITNET-2020 and
CSE-CIC-IDS-2018 Datasets.

| Classifiers | Dataset | Accuracy | Precision | Recall | F1-Score |
|---|---|---|---|---|---|
| RF | LITNET-2020 | 1.000000000 | 1.000000000 | 1.000000000 | 1.000000000 |
| RF | CSE-CIC-IDS-2018 | 0.990513682 | 0.96084231 | 0.859125805 | 0.893265813 |
| DT | LITNET-2020 | 1.000000000 | 1.000000000 | 1.000000000 | 1.000000000 |
| DT | CSE-CIC-IDS-2018 | 0.996709474 | 0.974639471 | 0.975416011 | 0.975026588 |
| XGboost | LITNET-2020 | 1.000000000 | 1.000000000 | 1.000000000 | 1.000000000 |
| XGboost | CSE-CIC-IDS-2018 | 0.992428337 | 0.958616558 | 0.904445321 | 0.927667115 |
| CNNs | LITNET-2020 | 0.989821970 | 0.983374832 | 0.989821992 | 0.986295708 |
| CNNs | CSE-CIC-IDS-2018 | 0.887517154 | 0.787686741 | 0.887517178 | 0.834627361 |
| DNNs | LITNET-2020 | 1.000000000 | 1.000000000 | 1.000000000 | 1.000000000 |
| DNNs | CSE-CIC-IDS-2018 | 0.887517154 | 0.787686741 | 0.887517178 | 0.834627361 |
| MLP | LITNET-2020 | 1.000000000 | 1.000000000 | 1.000000000 | 1.000000000 |
| MLP | CSE-CIC-IDS-2018 | 0.887517178 | 0.787686741 | 0.887517178 | 0.834627361 |
| RNNs | LITNET-2020 | 0.997327268 | 0.995994388 | 0.997327289 | 0.996437922 |
| RNNs | CSE-CIC-IDS-2018 | 0.887746000 | 0.861307000 | 0.887746000 | 0.835293000 |

Table 9 shows the performance metrics for multiple ML classifiers used on two different datasets LITNET-2020 and CSE-CIC-IDS-2018. The evaluation criteria include Accuracy, Precision, Recall, and F1-Score, which provide a complete view of each classifier's efficacy. For the LITNET-2020 dataset, the RF, DT, XGboost, DNNs, and MLP classifiers all received perfect scores (1.000000000), showing flawless prediction ability. The CNNs and RNNs classifiers likewise performed admirably, but somewhat below perfection, with RNNs exhibiting a modest fall in metrics compared to the others. When these classifiers were applied to the CSE-CIC-IDS-2018 dataset, most models showed a significant fall in performance metrics, with the exception of the DT and XGboost classifiers, which maintained high scores comparable to those attained on the LITNET-2020 dataset. On the CSE-CIC-IDS-2018 dataset, the CNNs, DNNs, MLP, and RNNs classifiers performed significantly worse than LITNET-2020 in terms of accuracy and F1-Score. This comparative analysis demonstrates the variability in classifier performance across datasets, emphasizing the importance of dataset characteristics and the potential need for model adjustment or selection based on the unique problems given by each dataset. While some classifiers remain stable across both datasets, others may require modifications to improve performance for certain types of network traffic or attack detection scenarios.

Table 10. Average Performance Metrics of ML Classifiers on the LITNET-2020 and CSE-CIC-IDS-2018 datasets.

| Classifiers | Accuracy | Precision | Recall | F1-Score | Average |
|---|---|---|---|---|---|
| RF | 0.995256841 | 0.980421153 | 0.929562903 | 0.946632907 | 0.962968451 |
| DT | 0.998354737 | 0.987319736 | 0.987708006 | 0.987513294 | 0.990223943 |
| XGboost | 0.996214168 | 0.979308279 | 0.952222661 | 0.963833558 | 0.972894666 |
| CNNs | 0.938669562 | 0.885530786 | 0.938669585 | 0.910461534 | 0.918332867 |
| DNNs | 0.943758577 | 0.893843370 | 0.943758589 | 0.917313681 | 0.924668554 |
| MLP | 0.943758589 | 0.893843370 | 0.943758589 | 0.917313681 | 0.924668557 |
| RNNs | 0.942536634 | 0.928650694 | 0.942536645 | 0.915865461 | 0.932397358 |

Table 10 illustrates the average performance of several ML classifiers across two cybersecurity datasets. Accuracy, Precision, Recall, F1-Score, and an overall Average score represent the classifiers' ability to correctly identify security threats. Table 13 shows the average performance of various ML classifiers on two cybersecurity datasets. Accuracy, Precision, Recall, F1-Score, and an overall Average score measure the classifiers' ability to correctly identify security threats. The RF classifier performs admirably, achieving near-perfect average accuracy and good scores across all criteria. The DT and XGBoost classifiers also produce excellent results, with the DT achieving the greatest average F1-Score and Average metrics, indicating a well-balanced performance in terms of both precision and recall. In comparison, CNNs, DNNs, and MLP classifiers have lower scores, notably in precision, which has an impact on their overall F1-Score and average performance. RNNs outperforms CNNs, DNNs, and MLP in terms of balance and precision, although it still lags behind tree-based classification algorithms. The table shows a wide range of performance, with tree-based classifiers beating neural network-based classifiers on these particular datasets. To better comprehend model performance evaluation metrics, the researcher presented the data in the form of a bar chart, as illustrated in Figure 3.

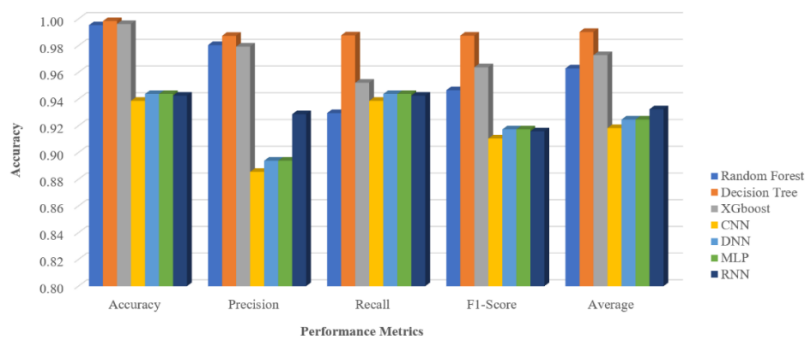

Figure 3. Comparative Bar Chart of Average ML Classifier Performance on LITNET-2020 and CSE-CIC-IDS-2018 Datasets.

The analysis in Tables 7-10 shows that the DT classifier outperforms alternative models on both the CSE-CIC-IDS-2018 and LITNET-2020 datasets. Based on this discovery, the researcher decided to improve the DT model by modifying three critical parameters max_depth, min_samples_split, and min_samples_leaf. To improve the model's performance further, EPSO was used. This optimization yielded the following optimal parameter settings for the DT: ccp_alpha=0.0,class_weight=None,

criterion='gini',max_depth=31,max_features=None,max_leaf_nodes=None,

in_impurity_decrease=0.0,min_samples_split=7,min_samples_leaf=1,

min_weight_fraction_leaf=0.0, and a random_state of 42, with the splitter set to 'best'. The favorable impact of these EPSO-facilitated tweaks is detailed in the following tables 14 and figures 4, which show that the DT model performs better now.

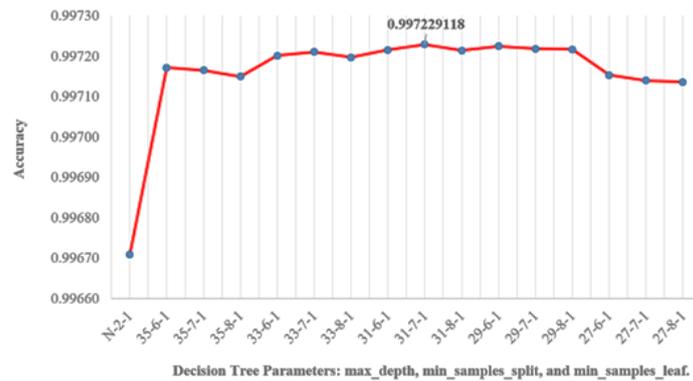

Figure 4. EPSO Driven Accuracy Improvements for DT on CSE-CIC-IDS-2018 Dataset.

The figure depicts the performance increase of a DT classifier applied to the CSE-CIC-IDS-2018 dataset after modifying its parameters using EPSO. The X-axis displays several parameter combinations for max_depth, min_samples_split, and min_samples_leaf. The Y axis represents the classifier's accuracy. Starting with the usual DT settings (max_depth is infinite, min_samples_split is 2, and min_samples_leaf is 1), there is a significant improvement in accuracy. This improvement reaches its pinnacle when the parameters are set to max_depth=31, min_samples_split=7, and min_samples_leaf=1, indicating the most successful parameter configuration optimized by EPSO, as indicated by a peak accuracy of around 0.997229118. Beyond this peak, accuracy levels off, meaning that making more changes to the parameters has little effect on improving accuracy. This high point is the most efficient iteration of the DT model, which was optimized using EPSO for the maximum predicted accuracy.

Table 11. Improving ML Classifier Performance with EPSO on CSE-CIC-IDS-2018 Dataset.

| Classifiers | Accuracy | Precision | Recall | F1-Score |
|---|---|---|---|---|
| RF | 0.990513682 | 0.960842310 | 0.859125805 | 0.893265813 |
| DT | 0.996709474 | 0.974639471 | 0.975416011 | 0.975026588 |
| EPSO DT | 0.997229118 | 0.983196382 | 0.967512492 | 0.974541397 |
| XGboost | 0.992428337 | 0.958616558 | 0.904445321 | 0.927667115 |
| CNNs | 0.887517154 | 0.787686741 | 0.887517178 | 0.834627361 |
| DNNs | 0.887517154 | 0.787686741 | 0.887517178 | 0.834627361 |
| MLP | 0.887517178 | 0.787686741 | 0.887517178 | 0.834627361 |
| RNNs | 0.887746000 | 0.861307000 | 0.887746000 | 0.835293000 |

Table 11 compares the performance of various classifiers on the CSE-CIC-IDS-2018 dataset, focusing on the DT and its upgraded variant utilizing upgraded Particle Swarm Optimization (EPSO). The typical DT is highly efficient, with an F1-Score of 97.50% and an accuracy of nearly 99.67%. However, applying EPSO to the DT improves its metrics greatly, increasing accuracy to around 99.72% and F1-Score to 97.45%. This enhancement highlights EPSO's importance in optimizing the classifier's parameters, hence improving predicted accuracy and reliability in detecting network intrusions. The EPSO-enhanced DT is the most effective model, demonstrating the promise of optimization techniques for improving ML tools for cybersecurity applications.

Table 12. Average Performance Metrics for EPSO-Tuned ML Classifiers on the LITNET-2020 and CSE-CIC-IDS-2018 datasets.

| Classifiers | Accuracy | Precision | Recall | F1-Score | Average |
| --- | --- | --- | --- | --- | --- |
| RF | 0.995256841 | 0.980421153 | 0.929562903 | 0.946632907 | 0.962968451 |
| DT | 0.998354737 | 0.987319736 | 0.987708006 | 0.987513294 | 0.990223943 |
| EPSO DT | 0.998614559 | 0.991598191 | 0.983756246 | 0.987270699 | 0.990309924 |
| XGboost | 0.996214168 | 0.979308279 | 0.952222661 | 0.963833558 | 0.972894666 |
| CNNs | 0.938669562 | 0.885530786 | 0.938669585 | 0.910461534 | 0.918332867 |
| DNNs | 0.943758577 | 0.893843370 | 0.943758589 | 0.917313681 | 0.924668554 |
| MLP | 0.943758589 | 0.893843370 | 0.943758589 | 0.917313681 | 0.924668557 |
| RNNs | 0.942536634 | 0.928650694 | 0.942536645 | 0.915865461 | 0.932397358 |

Table 12 summarizes the average performance metrics of ML classifiers optimized using EPSO on the LITNET-2020 and CSE-CIC-IDS-2018 datasets. Accuracy, Precision, Recall, and F1-Score are among the metrics used, with an overall average score for each classifier calculated as well. The EPSO-tuned DT performs best, with virtually flawless scores across all criteria, followed by the regular DT and RF classifiers, demonstrating their excellent efficacy in classification tasks. XGBoost also performs well, though with slightly lower averages. DLmodels, such as CNNs, DNNs, MLP, and RNNs, perform well but with lower average scores. The findings highlight EPSO's ability to refine models and offer precise and accurate forecast outcomes, particularly in difficult tasks like intrusion detection

Figure 5 shows a radar chart that compares the performance of numerous ML classifiers on the CSE-CIC-IDS-2018 and LITNET-2020 datasets using measures such as accuracy, precision, recall, F1-Score, and overall average score. The EPSO DT is featured as one of the classifiers, and its performance is demonstrated to be superior in nearly all criteria. This model, which was optimized using EPSO, stands out for its near-perfect accuracy of around 0.9986. Its Precision, which assesses the accuracy of positive predictions, is likewise quite high, nearly approaching 1.0. The Recall, which measures how well the model detects all relevant instances, is also quite high, implying that the EPSO DT excels at finding true positives. The F1-Score, which is the harmonic mean of Precision and Recall, is likewise approaching 0.99, indicating that the model performs well in both Precision and Recall. Finally, the Average score, which is presumably the mean of these metrics, is exceptionally high, demonstrating the classifier's overall robustness. In summary, the EPSO DT performs remarkably well across all tested measures, illustrating the value of employing EPSO to optimize DT in ML applications, particularly intrusion detection with the CSE-CIC-IDS-2018 dataset. The optimization process resulted in a model that is both accurate and trustworthy, with few trade-offs between identifying as many true positives as feasible and keeping a low false positive rate. To increase understanding of model performance evaluation criteria, the researcher displayed the data as a radar graph, as illustrated in Figure 5.

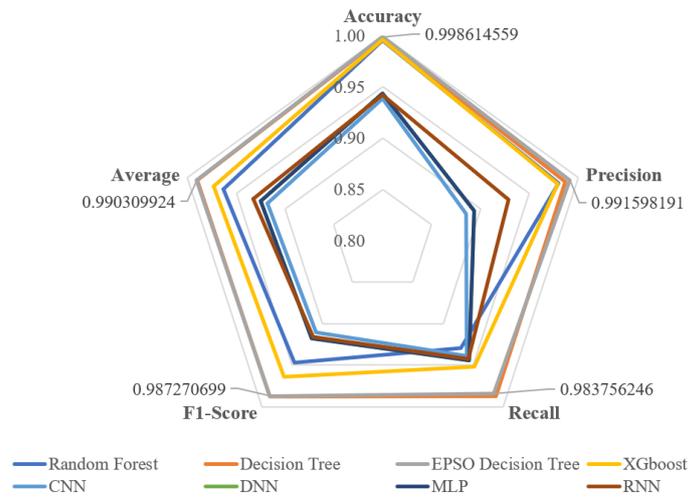

Figure 5. Comparative Radar Chart of ML Classifier Performances on
CSE-CIC-IDS-2018 and LITNET-2020 Datasets.

In the discussion of experimental results, we look at how ML classifiers performed after applying EPSO to two cybersecurity datasets CSE-CIC-IDS-2018 and LITNET-2020. The data was rigorously prepared for categorization by performing stages such as cleaning and normalization. The data was split into training and testing sets and processed with ML methods such as DT, RF, and XGBoost, as well as DL models such as CNNs, RNNs, DNNs, and MLPs. The classifiers were then compared on conventional criteria such as accuracy, precision, recall, and F1-Score, with the DT model consistently outperforming the others. The investigation found that the DT, when fine-tuned with EPSO to optimize important parameters, performed exceptionally well, achieving near-perfect accuracy and precision. This improved DT model proved to be the most effective, demonstrating EPSO's powerful impact on classifier performance. A radar chart comparison revealed that the EPSO-optimized DT outperformed all other measures, validating its efficacy in intrusion detection tasks on the relevant datasets. The optimization resulted in a classifier that not only reliably identifies threats but also has a low false positive rate, effectively balancing accuracy with recall. The findings hint at a promising approach for future advances in IDS, with EPSO emerging as a useful tool for improving ML models to attain high levels of prediction accuracy and reliability.

Table 13. Comparing Our Intrusion Detection Model to Others Using
CSE-CIC-IDS-2018 and LITNET-2020 Datasets.

| Study | Classifiers/Datasets | Accuracy | Precision | Recall | F1-Score |
|---|---|---|---|---|---|
| S. Songma, T. Sathuphan, and T. Pamutha. [3] | XGBoost-PCA11 / CSE-CIC-IDS-2018 | 0.997706000 | 0.920757000 | 0.988790000 | 0.949388000 |
| J. Kim, Y. Shin, and E. Choi, 2019 [38] | CNNs / CSE-CIC-IDS-2018 | 0.960000000 | n/a | n/a | n/a |
| M. A. Khan, 2021 [39] | HCRNNIDS / LITNET-2020 | 0.977500000 | 0.976000000 | n/a | n/a |
| L. Yang, J. Li, L. Yin, Z. Sun, Y. Zhao, and Z. Li, 2020 [16] | CDBN-based / LITNET-2020 | 0.974000000 | 0.966000000 | 0.976000000 | 0.971000000 |
| V. Dutta, M. Choraś, M. Pawlicki, and R. Kozik, 2020 [4] | DNNs / LITNET-2020 | 0.997000000 | n/a | n/a | n/a |

| V. Dutta, M. Choraś, M. Pawlicki, and R. Kozik, 2020 [4] | LSTM / LITNET-2020 | 0.991000000 | n/a | n/a | n/a |
|---|---|---|---|---|---|
| **Our Model** | EPSO-DT / CSE-CIC-IDS-2018 and LITNET-2020 | 0.998614559 | 0.991598191 | 0.983756246 | 0.987270699 |

Table 13 compares alternative intrusion detection models using the CSE-CIC-IDS-2018 and LITNET-2020 datasets, with a focus on key parameters including accuracy, precision, recall, and F1-Score. Our model, which uses an EPSO DT, stands out for its superior performance across both datasets. It has an excellent accuracy of around 99.86%, precision of 99.16%, recall of 98.38%, and F1-Score of 98.73%. This shows a high level of accuracy in identifying both actual and potential intrusions while avoiding false positives and negatives. The comparison highlights the efficacy of our EPSO DT model in improving the reliability and security of NIDS, establishing it as a highly competitive strategy in the field of cybersecurity.

## 7. CONCLUSION

In the conclusion portion of the work, we discuss the extensive testing and analysis of ML classifiers utilizing EPSO on two significant cybersecurity datasets CSE-CIC-IDS-2018 and LITNET-2020. The data was meticulously prepared, including cleaning, normalization, and partitioning into training and testing sets. The data was analyzed using a variety of ML approaches, including DT, RF, and XGBoost, as well as DL models such as CNNs, RNNs, DNNs, and MLPs. Each model was thoroughly tested with measurement metrics such as accuracy, precision, recall, and F1-Score. The DT model performed well, especially after EPSO fine-tuning, with nearly perfect scores across all criteria, suggesting its ability to appropriately identify data. The EPSO-optimized DT outperformed all other classifiers in radar chart comparisons, demonstrating its ability to identify network intrusions.

This extended research demonstrates EPSO's effectiveness as a potent optimization tool for improving the performance of ML classifiers in cybersecurity applications. The study concluded that with thorough preparation and optimization utilizing approaches such as EPSO, ML models can attain high levels of precision and dependability, which are critical for effective IDS. The findings provide useful insights for future cybersecurity advancements, underlining the importance of continuously improving analytical models to keep up with the changing nature of network threats.

Future research should focus on improving optimization approaches beyond EPSO, validating model generalizability across several datasets, and implementing models in real-time contexts. Exploring hybrid techniques and increasing resilience to adversarial attacks will improve IDS efficacy. This study's limitations include potential overfitting and the need for model validation across changing network conditions, necessitating further research into scalability, efficiency, and flexibility to emerging threats, particularly in IoT and edge computing contexts.

## CONFLICTS OF INTEREST

The authors declare no conflict of interest.

## ACKNOWLEDGMENTS

Our heartfelt gratitude goes to the team at Suan Dusit University for their invaluable insights and continuous support throughout this research. We are also appreciative to Suan Dusit University for providing crucial computer system resources that were critical in the execution and analysis of our study.

## AUTHORS

**Surasit Songma** his Ph.D. in Information Technology from Rangsit University. Currently he holds the position of Assistant Professor in the Cyber Security program at the Faculty of Science and Technology, Suan Dusit University. His current research interests include intrusion detection systems, network security, big data analytics, and machine learning. He has published several papers in peer-reviewed journals and has actively participated in various international conferences. He is a dedicated researcher with a passion for advancing the field of information technology through his work. He can contacted at email: surasit_son@dusit.ac.th

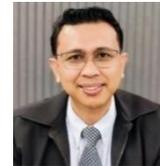

**Watcharakorn Netharn\*** his M.Sc. in Telecommunication and Computer Networks for Rangsit University. Currently he holds the position of Assistant Professor in the Information Technology program at the Faculty of Science and Technology, Suan Dusit University. His current research interest include satellite, network and data science. He is the *corresponding author for this article and can be contacted at watcharakorn_net@dusit.ac.th.

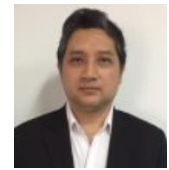

**Siriluck Lorpunmanee** his PhD in Computer Science from the University Technology of Malaysia. Currently, he holds the position of Assistant Professor in the Computer Science program at Suan Dusit University in Thailand. His current research interests include Artificial Intelligence Algorithm, Simulation modelling, and Grid computing. He has published several papers in peer-reviewed journals and has actively participated in various international conferences. He is a dedicated researcher with a passion for advancing the field of computer science through his work. He can contacted at email: siriluck_lor@dusit.ac.th

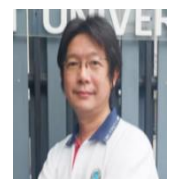